\documentclass[11pt]{article}
\usepackage{graphics}
\usepackage{paspconf}

\font\sit   = cmti9

\def\ie   {i.e.\ }
\def\iec  {i.e., }

\def\egc  {e.g., }
\def\etal {et al.\ }

\def\r    {\reference}       
\def\oi   {{\it (i)\/}~}
\def\ii   {{\it (ii)\/}~}
\def\iii  {{\it (iii)\/}~}
\def\iv   {{\it (iv)\/}~}
\def\v    {{\it (v)\/}~}

\def\parano    {\par \noindent}

\def\newpage{\vfill\eject} 
\def\lappeq{\lower 0.6ex\hbox{$ \buildrel < \over \sim$}~} 
\def\gappeq{\lower 0.6ex\hbox{$ \buildrel > \over \sim$}~} 
\def\md  {\lower 0.5ex\hbox{$\; \buildrel \times \over \div \;$} 
            \allowbreak}   
\def\e {$\;\>$}            
\def\x#1{$10^{#1}$}             
\def \prop   {$\propto\;$}
\def\ldots {...}                       
\def \bdot#1  {$\bf\dot{\rm #1}$}      
\def\hsla {{\raise1pt\hbox{\footnotesize /}}} 
\def\dBedt     {$d/dt\,{\rm Be}$}  
\def\dBedtISM  {$d/dt\,{\rm Be_{\rm ism}}$}  
\def\dBedtFAST {$d/dt\,{\rm Be_{\rm fast}}$}  
\def \alph {$\alpha\ $}

\def \alpho {$\alpha$}

\def \gammo {$\gamma$}

\def \tcheco#1 {$\check {\rm{#1}}$}
\def\ns      {nucleosynthesis\ }

\def\nc      {nucleosynthetic\ }
\def\Nc      {Nucleosynthetic\ }
\def\nso     {nucleosynthesis}

\def\SN      {SN\ae\ }
\def\SNII    {SN\,II's\ }
\def\SNIa    {SN\,Ia's\ }
\def\SNo     {SN\ae}
\def\SNIIo   {SN\,II's}
\def\SNIao   {SN\,Ia's}

\def\prefacc {preferential acceleration\ }

\def\cr      {cosmic ray\ }

\def\crs     {cosmic rays\ }

\def\CRs     {Cosmic Rays\ }

\def\crso    {cosmic rays}
\def\Crso    {Cosmic rays}

\def\sb      {superbubble\ }

\def\sbs     {superbubbles\ }

\def\sbso    {superbubbles}

\def\sol     {$_{\odot}$\ }
\def\Msol    {$M_{\odot}$\ }

\def\Msolo   {$M_{\odot}$}

\def\teco    {$\tau_{\rm ec}$}

\def\Ben       {$^{9}$Be\ }

\def\Boe       {$^{11}$B\ }
\def\Ct        {$^{12}$C\ }

\def\Crat      {$^{13}$C/$^{12}$C\ }

\def\Os        {$^{16}$O\ }

\def\Net       {$^{20}$Ne\ } 
\def\Nett      {$^{22}$Ne\ }
\def\Nerat     {$^{22}$Ne/$^{20}$Ne\ }

\def\Mnffour   {$^{54}$Mn\ }

\def\Fefsix    {$^{56}$Fe\ }

\def\Cofn      {$^{59}$Co\ }

\def\Nifn      {$^{59}$Ni\ }
\def\Nis       {$^{60}$Ni\ }

\def\Hefo      {$^{4}$He}

\def\Lisxo     {$^{6}$Li}

\def\Cto       {$^{12}$C}

\def\Neto      {$^{20}$Ne}
\def\Netto     {$^{22}$Ne}

\def\Feffouro  {$^{54}$Fe}

\def\Fefso     {$^{56}$Fe}
\def\Fefseveno {$^{57}$Fe}

\def\Cofno     {$^{59}$Co}
\def\Nifso     {$^{56}$Ni}
\def\Nifno     {$^{59}$Ni}

\def\eqb{\begin{equation}}
\def\eqe{\end{equation}}

\def\ie{i.e.\ }
\def\iec{i.e., }

\def\egc{e.g., }
\def\etal{et al.\ }
\newcount\listno
\def\list{\global\advance \listno by 1 {(\the\listno) }}

\newcount\eqnno
\newcount\eqnnoA
\def\eqnum{\global\advance \eqnno by 1 \eqno{(\the\eqnno)}}
\def\eqnumA{\global\advance \eqnnoA by 1 \eqno{(A\the\eqnnoA)}}
\def\eqnumal{\global\advance \eqnno by 1   (\the\eqnno)}
\def\label#1{\global\advance \eqnno by 1\xdef#1{\the\eqnno }
    \global\advance \eqnno by -1}
\def\labelA#1{\global\advance \eqnnoA by 1\xdef#1{\the\eqnnoA }
    \global\advance \eqnnoA by -1}
\def\AA {A\&A}      
\newcount\Inum
\newcount\IInum
\newcount\IIInum
\newcount\IVnum
\def\I{\global\multiply\IInum by 0 \global\multiply\IIInum by 0
            \global\multiply\IVnum by 0 \global\advance \Inum by 1
            {\the\Inum. }}
\def\II{\global\multiply\IIInum by 0\global\multiply\IVnum by 0
       \global\advance \IInum by 1 {\the\Inum.\the\IInum. }}
\def\III{\global\multiply\IVnum by 0\global\advance \IIInum by 1
            {\the\Inum.\the\IInum.\the\IIInum.}}
\def\IV{\global\advance \IVnum by 1
            {\the\IVnum.}}
\divide \count1 by 60
\def\hours{\number\count1}
\multiply\count2 by -60
\advance \count3 by \count2
\def\minutes{\number\count3}
\def\timeexplicit{\hours:\minutes}
\def\today{\rm{\space\space\space\space\space\number\day\space
           \ifcase\month\or January\or February\or March\or April\or May\or June
           \or July\or August\or September\or October\or November\or December\fi 
           \space\number\year\space\space\number\timeexplicit}}                         
\begin{document}



\title{\vskip-5mm The Origin of Present Day Cosmic Rays: \\
Fresh SN Ejecta or Interstellar Medium Material ? \\
I$\;\>$Cosmic Ray Composition and SN Nucleosynthesis. \\
\vskip 0.2truecm
A Conflict with the Early Galactic Evolution of Be ? *}

\author{\vskip-6mm Jean-Paul Meyer}

\affil{\vskip-4mm Service d'Astrophysique, DAPNIA, CEA/Saclay, \\ 91191
Gif-sur-Yvette, France \\ meyer@hep.saclay.cea.fr}

\author{\vskip-2mm Donald C. Ellison}

\affil{\vskip-1mm Department of Physics, North Carolina State University, \\ Box 8202,
Raleigh, NC 27695, USA \\ don\_ellison@ncsu.edu}

\begin{abstract}
We show that the composition of {\it present day\/} \crs is 
inconsistent with a significant acceleration of SN ejecta material 
(even if a preferential acceleration of ejecta grain material is 
assumed).
Current \crs must result essentially from the acceleration of 
interstellar gas and grain material, with a ``solar mix" composition, 
plus of circumstellar material (Wolf-Rayet wind \Netto- 
and \Cto-rich material).
The \cr source composition derived from observations,
indeed, shows {\it no\/} anomaly related to SN \nso.
Specifically:
\oi The \cr source FeNi/MgSiCa ratios have precisely the solar mix 
values, while FeNi are predominantly synthesized in \SNIao, and 
MgSiCa in \SNIIo.
To be understood in terms of an acceleration of SN ejecta, this would 
require tight conditions on the acceleration efficiencies of the 
ejecta of the various \SNIa and \SNII of all masses.
\ii The {\it lack\/} of a deficiency of the main-s-process
elements, not synthesized in {\it any\/} SN, relative to all elements made
in \SNo, is clearly inconsistent with a significant acceleration of
SN ejecta material.
\iii With the exception of the \Nett and related \Ct excesses,
suggesting the acceleration of Wolf-Rayet wind material, all
determined \cr isotope ratios are consistent with solar mix.
\iv The absence of \Nifn in \crs implies that the time delay between
the SN \ns of Fe peak nuclei and their acceleration is $\gappeq
10^{5}$ yr.
\v As discussed in Ellison \& Meyer, this volume, the physics of SNR's
and of \cr shock acceleration implies that the acceleration of {\it
interior\/} ejecta material is insignificant, as compared to that of
interstellar and/or circumstellar material {\it outside\/} the
forward shock.
Predominant acceleration of {\it current\/} \crs out of \sb material
also seems implausible.
\parano These conclusions regarding {\it current\/} \crs do {\it not\/}
necessarily conflict with the linear evolution of Be/H {\it in the
early Galaxy}.
With the near absence of heavy elements in the early Galactic ISM, 
indeed, the acceleration of even a {\it minute\/} amount of freshly 
processed material in the early 
%
\begin{minipage}{12.24cm}  
\vskip 0.15truecm 
\parano  \leftskip=-11mm -----------------------\vskip -0.2truecm
{\sit \parano \leftskip=-11mm {{\rm*} Invited paper, Workshop on "LiBeB, 
Cosmic Rays and Gamma-Ray Line Astronomy",\break\vskip-5mm 
IAP, Paris, 9-11 December 1998, 
M.\ Cass\'e, K.\ Olive, R.\ Ramaty, \& E.\ Vangioni-Flam eds.,\break\vskip-5mm
ASP Conf.\ Series 
(Astr.\ Soc.\ of the Pacific, 1999), in press.\par}}
\end{minipage}
\newpage

\noindent Galaxy (SN ejecta ?  Wolf-Rayet winds ?  superbubbles ?)  
must have played a dominant role for the generation of Be from C and 
O. The ``Be indicator" is, indeed, blind to a possibly dominant early 
Galactic \cr component originating in the ISM then composed of 
virtually pure H and He.
\end{abstract}


\section{Introduction}                                  

\subsection{The Early Galactic Evolution of the Be Abundance}

\subsubsection{1.1.1.  The situation, based on the conventional determinations}
~ \vskip 0.0001truecm
\parano {\leftskip=1.15cm {\it of the evolution of the Galactic O/H ratio} \par}
\vskip 0.3truecm

\parano Observations indicate that the Galactic Be/H and B/H ratios 
increase close to linearly with Fe/H, at least in the early Galaxy, 
for [Fe/H] between --~3 and --~0.5
%
\footnote{As usual [A/B] denotes log$_{10}$(A/B) -- log$_{10}$(A/B)\sol,
where A and B are the abundances of two elements.}
%
(\egc Molaro \etal 1997; Duncan \etal 1997; Garc\'\i a L\'opez \etal 1998;
Vangioni-Flam \etal 1998a for a review).
Now, the significant correlation is {\it not\/} Be/H vs.\ Fe/H, but
rather Be/H vs.\ O/H, since O is, with C, the principal progenitor
of Be
%
\footnote{From here on, we will consider only Be.  The Be and B 
origins and histories share many similarities.
But, contrary to Be, which can be made {\it only\/} by proton and 
\alpho-particle induced spallation of heavier nuclei in space, \Boe 
can be made by neutrino induced spallation in Type-II \SNo, as well 
(Woosley \& Weaver 1995; Timmes \etal 1995).
In addition, the particle induced spallation yields of \Boe are 
particularly sensitive to the exact shape of the interacting 
particle spectrum at low energies (\egc Meneguzzi \&  Reeves 1975; 
Ramaty \etal 1997).
So, the interpretation of behavior of B is less straightforward than 
that of Be.}.

On the other hand, the conventional studies of the evolution of the
O/Fe ratio, largely based on [OI] forbidden line observations,
indicate a constancy of O/Fe at a value of $\sim2.5\,\times$ solar,
from [Fe/H] $\sim$ -- 1 down to -- 3 (Pagel \& Tautvai\tcheco{s}
ieni\.e 1995; McWilliam 1997).
The linear correlation Be\hsla H \prop Fe\hsla H thus translates into
a linear correlation with O\hsla H as well, \ie Be\hsla H \prop O\hsla H.
This implies that, in the early Galaxy, the Be production rate 
\dBedt\ was independent of the Galactic abundance of its major 
progenitor, O, \ie that Be is ``primary".

Now, Be can be synthesized only by high energy spallation of 
essentially CNO nuclei interacting with protons
%
\footnote{For legibility, we denote by ``interacting with protons" 
what should actually read ``interacting with protons and 
\alpho-particles", since the latter also play a significant role.
Note also that the role of N is small, compared to those of C and O.}.
%
There exists two channels for these interactions:
\oi the spallation of ISM\e CNO, at rest, by \cr protons, producing Be
at rest in the ISM (``Be$_{\rm ism}$"), and
\ii the spallation of fast \cr CNO's on ISM protons at rest, producing
fast Be nuclei (``Be$_{\rm fast}$").

Since the Be$_{\rm ism}$ component originates in the ISM\e CNO nuclei, 
its production rate \dBedtISM\ increases roughly as the Galactic O/H 
(\dBedtISM\ \prop O/H), so that the Be$_{\rm ism}$ component is purely 
``secondary" (Be$_{\rm ism}$/H \prop O/H$^{2}$)
%
\footnote{This assumes that the interacting \cr protons are not 
essentially {\it low energy\/} particles ($\sim$10's of MeV/nucleon) 
with a small stopping range, {\it confined\/} within regions {\it 
locally enriched\/} in heavy elements, such as superbubbles.
This possibility seems very unlikely, since superbubbles have very low 
densities, so that even low energy particles will mainly interact (and 
be stopped) in nearby dense ISM clouds, rather than within the 
enriched superbubble material itself (\egc Bykov 1995).  
This is all the more so, that only the central part of \sbs is 
actually enriched, the gas in most of their volume being dominated by 
material evaporated from nearby dense clouds (Higdon \etal 1998).}.
%
As for the Be$_{\rm fast}$ component, it has a primary or a secondary
character, according to whether the \crs has been accelerated out of
\oi the general ISM (\dBedtFAST\ \prop O/H~: secondary), or
\ii freshly processed material, such as SN ejecta or Wolf-Rayet star
wind material, in which the large CNO abundance is independent of
Galactic O/H (primary).
Therefore, only a Be production controlled by the spallation of \cr
CNO nuclei accelerated out of freshly processed material can account
for the primary character of the observed evolution of the early
Galactic Be/H ratio.
This implies, in particular, that, {\it in the early Galaxy\/}, most
\cr CNO nuclei originated, {\it not\/} in the ISM, but in fresh
sources of \nso, and most plausibly from SN\,II ejecta
%
\footnote{For legibility, we denote by ``SN\,II" all massive star \SN 
of Type II and Ib.}
%
(\egc Duncan \etal 1992; Feltzing \& Gustafsson 1994; Vangioni-Flam \& 
Cass\'e 1996; Ramaty \etal 1997,1998; Lingenfelter \etal 1998; 
Vangioni-Flam \etal 1998a; Higdon \etal 1998).

This is, actually, not a surprise.  In the early Galaxy, indeed, there
was very little CNO in the ISM, so that very little Be could originate
in the spallation of any material originating in it
(\iec both of ISM\e CNO, and of \cr CNO accelerated out of this ISM).
Therefore, any contribution, however minor, to \crs from freshly 
synthesized, CNO-enriched material could easily yield a dominant 
contribution to the Be production, and make Be evolve as a primary in 
the early Galaxy -- until a significant amount of CNO resided in the ISM 
({\S}~3).

%
%


\subsubsection{1.1.2.  The situation, based on the recent determinations}
~ \vskip 0.0001truecm
\parano {\leftskip=1.15cm {\it of the evolution of the Galactic O/H ratio 
                               (with OH molecular lines)} \par}
\vskip 0.3truecm

\parano Recent studies, based on OH molecular line observations, yield 
a different evolution of the Galactic O/Fe ratio with Fe/H:
O/Fe seems to continuously increase with decreasing Fe/H from [Fe/H]
$\sim$ 0 down to -- 3, with a slope of $\sim$~-- 0.35 (Israelian \etal
1998; Boesgaard \etal 1999).
If this is correct, the linear evolution Be/H $\propto$ Fe/H translates
into a new relationship Be/H $\propto$ O/H$^{1.35}$.  A more elaborate
treatment by Fields \& Olive (1998) yields Be/H $\propto$ O/H$^{1.3\
\rm to\ 1.8}$.
The validity of the new O abundance determinations based on OH lines
is currently a matter of debate (Vangioni-Flam \etal 1998b; Cayrel
1999)
%
\footnote {The new OH molecular line studies imply high O abundances
for the very early Galaxy, which might also pose energetic problems
for their production.
This behavior of O/H would also have to be understood along with the
behavior of the other \alpho-elements (``\alpho\,"), presumably synthesized
in the same Type-II \SNo, which apparently do not show a similar
increase of the ``\alpho\,"/Fe ratio in the [Fe/H] $\sim$ --1 to -- 4
range
(\egc Pagel \& Tautvai\tcheco{s} ieni\.e 1995; McWilliam 1997).}.
%

These new data on the Galactic O\hsla H thus imply a Be production 
rate \dBedt\ which increases with O\hsla H, but not as fast as \dBedt\ 
\prop O/H, as would be the case if Be were purely ``secondary" (Be/H 
\prop O/H$^{2}$).
If they are valid, we should have had, in the early Galaxy,
significant Be contributions from spallation of {\it both\/} \cr CNO
accelerated out of freshly processed material (primary), and of CNO
originating in the ISM (CNO at rest in the ISM, broken up by the \cr
protons, and/or \cr CNO nuclei accelerated out of the ISM material,
both secondary).

\subsection{The Current Galactic Cosmic Ray Source (GCRS) Material}

On the other hand, the observed composition of the {\it current\/}
\crs seems to imply an acceleration out of interstellar and/or
circumstellar material, with a preferential acceleration of the grain
material over the gas-phase ions, and {\it not\/} an acceleration of
SN ejecta (Meyer \etal 1997,1998; Ellison \etal 1997; Ellison \&
Meyer 1999).

The current \cr source composition is shown in Fig.~1, compared to
that of the Sun, versus mass $A$.  The elements are sorted according
to their volatility.
The refractory elements, locked in grains in the ISM, are found
globally enhanced relative to the volatile ones, which remain in the
gas-phase.
For the volatile elements, the enhancements increase with mass $A$
%
\footnote{Except for H, which has a high thermal speed possibly 
comparable to the viscous subshock speeds, C, which has a large 
Wolf-Rayet component and is partly locked in grains, and O, also 
partly locked in grains.}.
%
For the refractory elements, by contrast, there is only a very weak
increase of the enhancements with mass $A$, if any
%
\footnote{This is evident from the behavior (shown in Fig.~1) of those 
refractory elements whose \cr source abundance is accurately 
determined, \ie Mg, Al, Si, Ca, Fe, Ni and, to a lesser degree, Sr, Zr 
($A$~=~24 to $\sim$~90); see {\S}~2.3.  The absolute source 
abundances of the elements from Mo upward ($A$ \gappeq 95) are 
affected by large, partly systematic, uncertainties.}
%
(Meyer \etal 1997,1998).

%
%
\begin{figure}                   
\resizebox{\hsize}{!}{\includegraphics{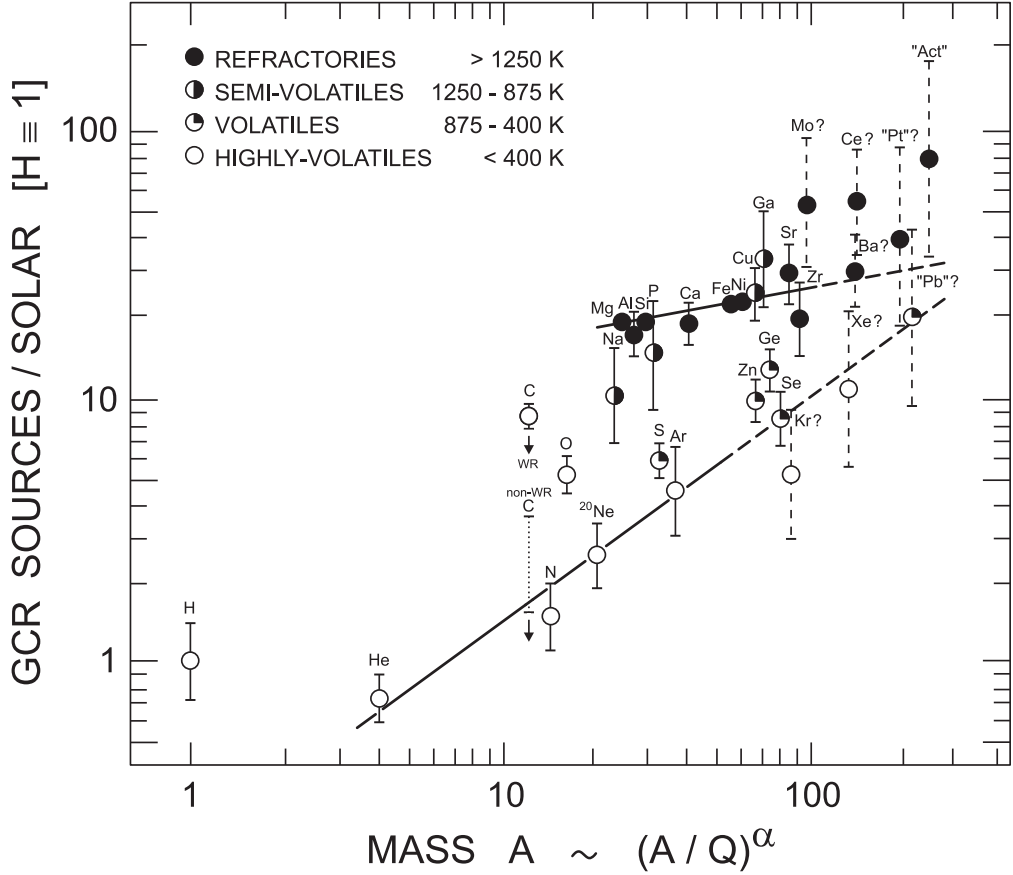}}
\caption{\small The Galactic Cosmic Ray Source (GCRS) to Solar 
abundance ratios in the $\sim$~GeV range, versus element mass $A$, 
with the elements sorted according to their volatility (after Meyer 
\etal 1998 and Westphal \etal 1998).
See text {\S}~1.2.
Normalized to H, at a given energy/nucleon.
The point for C is plotted as an upper limit, since its total source
abundance includes a specific \Ct contribution associated with a
\Netto-rich component presumably from WR star wind material;
we propose an estimate of the {\it non\/}-WR C source abundance
(which may still be an overestimate, since we did not consider any
preferential acceleration of C -- particularly locked in grains in the
C-rich WR wind material -- relative to \Nett in the WR component);
for Ne, we have plotted the \Net abundance.
``Pt", ``Pb", and ``Act" stand for $Z$~=~74--80, 81--83, and  90--92,
respectively, the latter being normalized to the relevant undecayed, {\it
proto\/}-solar abundances.
We have marked by a dashed error bar and a ``?"  sign those 
ultra-heavy elements whose source abundance {\it relative to Fe\/} is 
quite uncertain; but the relative abundances of {\it neighboring\/} 
elements (\egc the ``Pb"/``Pt", ``Act"/``Pt" ratios) is much better 
determined.
The lines roughly fitting the ``highly-volatile" and the
``refractory" element points are just to guide the eye; they are
represented solid in the well determined range, and extrapolated
dashed in the much less well established, very heavy element, range.
Note that these lines are {\it not} the predictions from the Ellison 
et al.\ (1997) model.  In fact, the Ellison et al.\ model does not 
predict a strict power law for the highly-volatile element 
enhancements, and matches the H and He observations better than the 
power law shown here.}
%
\end{figure}
%
%
%

These contrasting behaviors are simultaneously interpreted in terms of 
an acceleration of the external (interstellar and/or circumstellar) 
material traversed by the SNR shocks, which get smoothed by the 
backpressure of the accelerated particles which is always substantial 
if the acceleration is efficient.  This smoothed (i.e., nonlinear) 
shock acceleration leads to a higher acceleration efficiency for ions 
with higher mass-to-charge ratio $A/Q$, which ``see" a larger fraction 
of the entire shock velocity gap at each crossing of the shock.
For the volatile elements, the observed increase of the enhancements
with $A$ just reflects this increase of the acceleration efficiency
for ions with higher $A/Q$ ratio in the gas-phase.
On the other hand, dust grains are slightly charged, and should behave 
like ions with huge $A/Q$ ratios of order $\sim$~\x8.  As such, they 
are very efficiently accelerated up to $\sim$~0.1 MeV/nucleon, where 
the friction on the gas both cancels the acceleration and sputters off 
some \x{-4} of the grain mass.  Refractory element ions are thus 
injected at $\sim$~0.1 MeV/nucleon.  These then get further 
accelerated, with high efficiency, as individual ions.
Since in the {\it crucial\/}, early phases, the refractory elements
are accelerated, {\it not\/} as individual ions, but as constituents
of entire grains, their enhancements are expected to be roughly independent
of their mass, as observed
(Ellison \etal 1997; Ellison \& Meyer 1999).

Note that we have definite evidence for the presence of a 
nucleosynthetically peculiar component in \crso:
the observed \Nett excess by a factor of $\sim$~4.5 in the \cr sources.
This \Nett excess strongly suggests the presence of a Helium-burning
material component in \crso, most likely originating in WC-type
Wolf-Rayet star (WR) wind material.
An associated \Ct excess is expected, which has been assessed in Meyer
\etal (1997), and is, indeed, suggested by the recent isotopic
observations (see {\S}~2.4).
Fig.~1 therefore shows both the total cosmic ray C abundance and the
estimated {\it non\/}-WR component (which may still be an
overestimate, see caption).
Note that, in the above SNR shock acceleration scenario, the shocks
associated with the most massive \SN will naturally accelerate the
(external) pre-SN WR wind material, which is enriched in \Nett and \Ct
(Meyer \etal 1997).

\subsection{The puzzle -- Our Questioning}

With {\it current\/} \crso, some $\sim$~93~\% of the Be
produced is of secondary origin.
Consider, indeed, the two channels for Be production mentioned in
{\S}~1.1.1.
\oi The spallation of ISM\e CNO by \cr protons, Be$_{\rm ism}$;
this component is purely secondary.
\ii The spallation of fast \cr CNO's on ISM protons, Be$_{\rm fast}$;
this component, by contrast, is only $\sim$~64~\% secondary, and
$\sim$~36~\% primary, due to the large contribution of WR star wind
material to the \cr C
%
%
\footnote {The main progenitors of the Be$_{\rm fast}$ component are,
indeed, the \cr O and C nuclei.  
The role of N is small, and will be neglected in a first 
approximation.
According to the analysis of Meyer \etal (1997,1998), the \cr O 
originates entirely in the ISM (with some $\sim$~20~\% of the O locked 
in grains), so that O is a secondary progenitor.
As for C, {\it at least\/} $\sim$~75~\% -- so, say, 85~\% -- of its
source abundance originates in WC-type WR star wind material (see
Fig.~1 caption).
This WR\e C has been ultimately produced out of the stellar H (first
turned into \Hefo, and then into \Cto), and has therefore a primary
character (Meyer \etal 1997).
Now, in current $\sim$~GeV \crs propagating in the Galaxy, the C/O
ratio is $\sim$~1.09, with $\sim$~0.69 accelerated out of WR wind
material, $\sim$~0.12 accelerated out of the ISM, and $\sim$~0.28
resulting from, mainly O, spallation (Engelmann \etal 1990; Meyer
\etal 1998).
Further, the weighted ratio of the p- and \alpho-induced, high energy 
spallation cross sections for \Ben formation from \Ct and from \Os is 
$\sim$~1.22 (Ramaty \etal 1997).
With these figures, we find that altogether, about $\sim$~36~\% of the 
currently produced Be$_{\rm fast}$ is of primary origin.}.
%
%
Now, for standard \cr spectral shapes, this Be$_{\rm fast}$ component
makes up only some 20 \% of the total galactic Be, because most of the
formed Be nuclei escape the Galaxy or break up (Meneguzzi \& Reeves
1975)~
%
%
\footnote {This may, however, not be true if there exists in the ISM, 
a ``carrot" of intense \cr fluxes at such low energies ($\lappeq$~100 
MeV/nucleon) that they cannot be observed within the Solar cavity, 
being excluded by the interaction with the expanding Solar wind (\egc 
Meneguzzi \& Reeves 1975).}.
%
%
Altogether, for standard \cr spectra, only $\sim$~7~\% of the 
currently produced Be is of primary origin.

Recently, Ramaty \etal (1997,1998) and Lingenfelter \etal (1998) have
contended that the evolution of Be/H {\it in the early Galaxy\/}
requires that {\it current\/} \crs are accelerated out of fresh SN
ejecta.
Along this line, Lingenfelter \etal (1998) have argued that a
preferential acceleration of grain over gas-phase material {\it within
fresh SN ejecta\/} might account for the observed \cr composition and
for its acceleration -- as well as such an acceleration out of
interstellar and/or circumstellar material.

Here we wish to reexamine whether {\it current\/} \crs could be
accelerated out of fresh SN ejecta material.  We will conclude that
this seems impossible, in view of two types of difficulties:
\oi difficulties with the observed \cr composition, discussed in terms
of SN \ns in {\S}~2, and
\ii difficulties with the SNR physics and the shock acceleration of
ejecta material, discussed in a companion paper by  Ellison \&
Meyer (1999).

We will then ask ourselves whether these conclusions regarding {\it 
current\/} \crs necessarily conflict with the {\it early Galactic\/} 
evolution of Be/H, and conclude that it may {\it not\/}.  
Finally, we suggest possible ways out of this apparent contradiction 
({\S}~3).

Note that in the same spirit, but along a different line, Higdon \etal 
(1998) have later suggested that {\it current\/} \crs might be 
accelerated out of superbubble material enriched in fresh \ns 
products.  This also seems very difficult, both in view of the 
observed \cr composition ({\S}~3.2.3) and of the \sb physics (Ellison 
\& Meyer 1999).

\section{Is the Composition of Current Cosmic Rays Consistent with \\
         the Acceleration of Fresh SN Ejecta ?}         

\subsection{The \Nc Origins and Abundances
            \\ of the Cosmic Ray Elements}            

Fig.~2, like Fig.~1, shows the GCRS/Solar enhancements versus mass,
but with the elements sorted, this time, according to their \nc
origin
%
\footnote{In the context of \nso, it is important to recall the 
significance of the ``solar" or, better, ``solar mix" abundances.  The 
``solar mix" abundances result from the cumulated contributions of the 
\nc yields of many different types of stars (over a wide range of 
masses and life times) throughout the life of the Galaxy until the 
birth of the Sun, which has led to the solar and to the, roughly 
similar, current local ISM composition.}.
%
We have, nevertheless, still indicated whether an element is a
full-fledged refractory (closed symbols, or ``$\times$"), or not (open
symbols, for all more or less volatile elements).
The elements have been sorted into four types of \nc origins
%
\footnote{References for \nc origins, general: Anders \& Grevesse 
(1989); Wheeler \etal 1989; Edvardsson \etal (1993); Pagel \& 
Tautvai\tcheco{s} ieni\.e (1995); Woosley \& Weaver (1995); Timmes 
\etal (1995); McWilliam (1997); Arnould \& Takahashi (1999).}:

%
%
\begin{figure}[h]                   
\resizebox{\hsize}{!}{\includegraphics{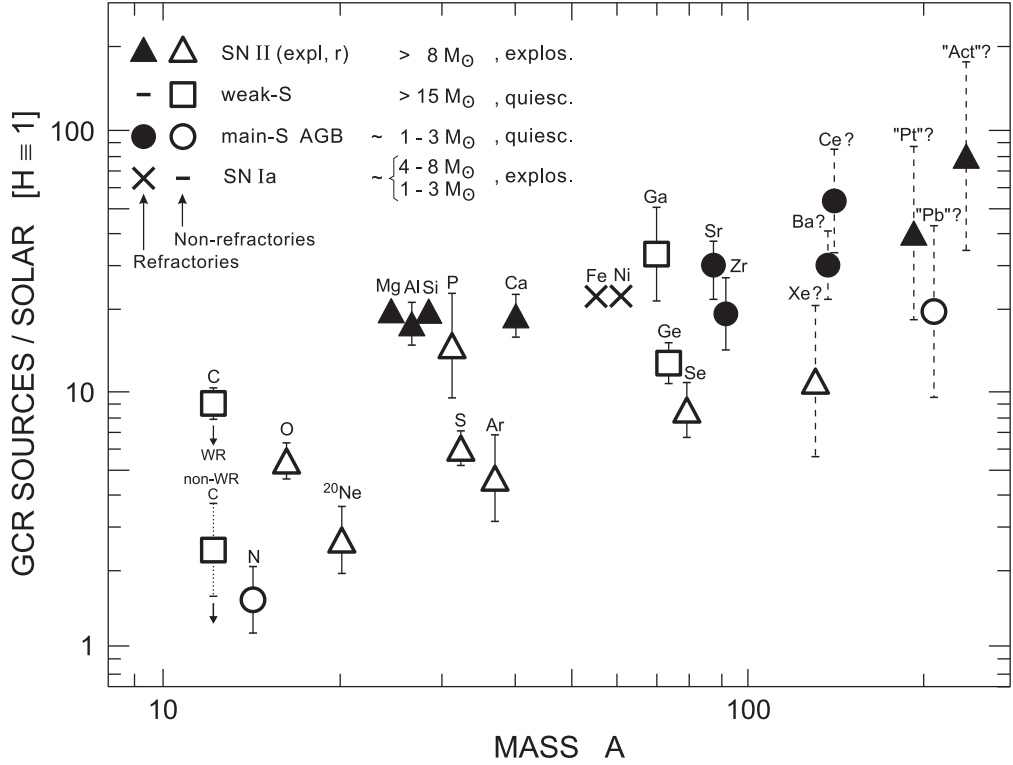}}
\caption{The same GCRS to Solar abundance ratios as in Fig.~1, with 
the elements sorted, this time, in terms of their \nc origin.  
Explanations in the text, {\S}~2.1.}
\end{figure}
%
%
%

\vskip 0.15truecm \parano
\oi {\it Explosive \ns in SN\,II\/}, \ie in massive stars with $> 8$ \Msol
initial mass
%
\footnote{References for SN\,II nucleosynthesis: Arnett (1995);
Woosley \& Weaver (1995); Timmes \etal (1995); Thielemann \etal
(1996); Nomoto \etal (1997).}.
%
This includes O, \Neto, Mg, Al, Si, P, S, Ar, Ca, and the predominantly
r-process elements Se, Xe, Pt-group, Actinides
%
\footnote{Pt-group: $Z$ = 74--80 , Pb-group: $Z$ = 81--83 , Actinides:
$Z$ = 90--92.}
%
(triangles).

\vskip 0.1truecm \parano
\ii {\it Quiescent weak-s-process\/} in massive stars with $> 15$
\Msol initial mass; the material is also ejected in the SN\,II
explosion, and possibly in stellar winds
%
\footnote{References for the s-process, general (weak and main): \egc
K\"appeler \etal (1989); Palme \& Beer (1993); Meyer (1994); Beer
\etal (1997); Gallino \etal (1998).}.
%
This group includes the predominantly s-elements with mass $A$ \lappeq
87, \ie Ga and Ge (squares).

\vskip 0.1truecm \parano
\iii {\it Quiescent main-s-process\/}, taking place in low mass, 1 --
3 \Msol stars during the AGB phase.
This group includes the predominantly s-elements with mass $A$ \gappeq
87, \ie Sr, Zr, Ba, Ce, Pb-group (circles)
%
\footnote{The limit between the predominance of the weak- and of the
main-s-process is, of course, not abrupt, and somewhat model
dependent.  For species with $A \leq 85$, the weak-s-process is
clearly predominant, while the main-s-process clearly dominates for
$A \geq 88$
(\egc K\"appeler \etal 1989; Beer \etal 1992; Baraffe \etal 1992;
Raiteri \etal 1993; Palme \& Beer 1993; Gallino \etal 1998).
Note that the solar Sr ($Z$ = 38), with essentially three isotopes
($^{86}$Sr,$^{87}$Sr,$^{88}$Sr), is 83 \% $^{88}$Sr (Anders \& Grevesse
1989).  So, Sr is essentially a main-s-process element.}.

{\vskip 0.1truecm \parano
\iv {\it Explosive \ns in SN\,Ia\/}, \ie in intermediate--lower mass
star binary systems having formed a white dwarf, with initial masses
of 4 -- 8 \Msol for the white dwarf progenitor and 1 -- 3 \Msol for
the companion
%
\footnote{References for SN\,Ia \ns: Thielemann \etal (1986);
Edvardsson \etal (1993); Yoshii \etal (1996); Kobayashi \etal (1998).
See discussion of the Fe,Co,Ni \ns sites in {\S}~2.3.1.}.
%
This group includes Fe and Ni (``$\times$" signs).
\vskip 0.1truecm

For C, both the total, presumably WR wind dominated, and the {\it
non\/}-WR \cr C points have been {\it tentatively\/} denoted by
squares,
indicative of a predominantly quiescent origin in massive stars, with
wind or explosive ejection (like the weak-s elements)
%
\footnote{References for C and N \ns: Wheeler \etal (1989); Pagel
(1992,1994); Andersson \& Edvardsson (1994); Timmes \etal (1995);
McWilliam (1997); Portinari \etal (1998); Gustafsson \etal (1999).}
%
%
\footnote{Regarding C, we have a special situation, in view of the
specific \cr \Ct component associated with the \Nett excess,
presumably accelerated out of WR wind material ({\S}~1.2).
The total \cr C abundance, probably dominated by this WR component,
has been denoted by a square in Fig.~2
(like weak-s elements, quiescent origin in massive stars, and wind or
explosive ejection).
As for the {\it non\/}-WR \cr C, it refers to the nucleosynthetic
origin of the general C abundance in the Galaxy, which is currently a
hot issue; there are probably significant contributions from stars of
very different masses; following Gustafsson \etal (1999), who
conclude that WR winds should also play a dominant role in the
synthesis of the Galactic C, we {\it tentatively\/} also plot the
non-WR C point with a square.}.
The N point has been denoted by a circle, since N seems predominantly,
though not entirely, made in low mass stars (like the main-s
elements).
But C and N are, anyhow, not important for the subsequent discussion.

In Fig.~2, we have omitted Na, Cu, and Zn, whose \nc origin isn't
firmly established (Edvardsson \etal 1993; Timmes \etal 1995;
McWilliam 1997), Mo and Kr, which have comparable s- and r-process
contributions, as well as H and He. 
%

\subsection{The \Neto/Mg and S,Ar/Si,Ca Ratios}      

The low GCRS \Neto/Mg and S,Ar/Si,Ca ratios cannot be explained in
terms of SN \nso. 
%
On the other hand, all these elements are believed to be essentially
synthesized in Type II \SNo, and their average relative abundances in
SN\,II ejecta should be similar to the solar mix or ISM ones.
Therefore, a \prefacc of grain material over gas ions may yield
similar GCRS ratios if applied to, either ISM material, or SN ejecta.
If it accounts for these ratios for an accelerated ISM material (Meyer
\etal 1997), it may as well account for them for an accelerated ejecta
material (Lingenfelter \etal 1998).

So, as noted by Lingenfelter \etal (1998), the low GCRS \Neto/Mg and 
S,Ar/Si,Ca ratios do not allow one to discriminate between an 
acceleration of ISM or SN ejecta material, provided that a \prefacc of 
grain over gas-phase material can actually work in both environments 
(discussion in Ellison \& Meyer 1999).

\subsection{The Refractory Elements of Various Nucleosynthetic Origins}

We now consider only the 12 refractory elements, whose \cr source
abundances and nucleosynthetic origin are reasonably well determined:
Mg, Al, Si, Fe, Ni, Sr, Zr, Ba, Ce, Pt-group, Actinides.

As shown in Fig.~2, they are all found to be in proportions close to 
the solar mix ones in the \cr sources: to within 20~\% for Mg, Al, Si, 
Fe, Ni, to within a factor of $\sim$~1.5 for Sr and Zr, and a factor 
of $\sim$~3 (upward) for Ba, Ce, Pt-group and Actinides (Meyer \etal 
1997,1998; Westphal \etal 1998)
%
\footnote {There are still possible systematic errors on the Ba, Ce, 
Pt-group and Actinide \cr source abundances {\it relative to\/} much 
lighter elements, such as Fe -- mainly due to our poor knowledge of 
the precise shape of the \cr interstellar pathlength distribution 
below $\sim$~1~g~cm$^{-2}$ (see Meyer \etal 1997).}.
%
This fact is very unlikely to be fortuitous, and strongly suggests
that no strong fractionation related to, either chemistry, or atomic
physics, or nucleosynthetic origin is at work among them
%
\footnote {There may be a {\it weak\/}, smooth increase of the GCRS
abundances with mass, among the refractory elements; see interpretation
in Meyer \etal (1997).}.

\vbox{
Now, these elements are synthesized in very different environments:
\vskip 0.15truecm \vskip 0.15truecm
\parano
Elements~~~~~~~~~~~~~~~~~~~~Mg,Si,Ca~~~~~~~~~~~~~~Fe,Ni~~~~~~~~~~~~Sr,Zr,Ba,Ce~~~~~~~~~Pt-gr,Act.
\vskip 0.15truecm
\parano made by
~~~~~~~~~~~~~expl.O,Si-burn.~~~~e-process~~~~~main-s-process~~~~~~r-process
\vskip 0.15truecm
\parano in ~~~~~~~~~~~~~~~~~~~~~~~~~~~SN II~~~~~~~~~~~~~~SN
Ia~~~~~~~~~~AGB-stars~~~~~~~~~~SN II
\vskip 0.15truecm
\parano with initial masses ~~~$>8$ \Msol~~~~~~~~~~$4 - 8$ \Msol~~~~~~~$1 -
3$ \Msol~~~~~~~~$>8$ \Msol
\parano ~~~~~~~~~~~~~~~~~~~~~~~~~~~~~~~~~~~~~~~~~~~~~~$\>$+ $1 - 3$ \Msol
\vskip 0.15truecm \vskip 0.15truecm
}

Nevertheless, as we have just seen, they are all found to have nearly 
solar proportions in \cr sources !

We now investigate more closely the significance of two ratios in the
GCRS: the Fe,Ni/Mg,Si,Ca and the main-s-elements/all-others ratios.

\subsubsection{2.3.1. The Fe,Ni/Mg,Si,Ca ratios} ~ \vskip 0.3truecm

\parano Virtually all the galactic Mg,Si,Ca is synthesized in \SNIIo.
By contrast, about $\sim$~70~\% of the Fe,Co,Ni is synthesized in
\SNIao, and only $\sim$~30~\% in \SNII
(\egc Edvardsson \etal 1993; Pagel \& Tautvai\tcheco{s} ieni\.e 1995;
Timmes \etal 1995; Yoshii \etal 1996)
%
\footnote{This estimate is mainly derived from the contrasted
evolutions of the galactic Fe/H and O,Mg,Si,Ca/H ratios, the latter
elements being all made in \SNIIo.
It may be noted that the {\it nominal\/} model of Woosley \& Weaver 
(1995) and Timmes \etal (1995) attributes a twice larger fraction of 
Fe to \SNIIo; but the Fe yield of \SNII is very dependent on 
the poorly known mass-cut; Timmes \etal (1995) have actually noted 
that a twice smaller Fe contribution of \SNII fits the galactic 
evolution data better, as well as the SN 1987A observations.
Note also that Co and Ni are, like Fe, predominantly made in \SNIao, 
as evidenced by the constancy of the galactic Co/Fe and Ni/Fe ratios 
versus Fe/H (Gratton \& Sneden 1991; Edvardsson \etal 1993; McWilliam 
1997).}.
%
Nevertheless, the \cr source Fe,Ni/Mg,Si,Ca ratios are equal to those 
in the solar mix, to within 20~\%.

Lingenfelter \etal (1998) have shown that an average of the calculated 
yields of the various types of \SN over the Initial Mass Function 
(IMF) leads to Fe,Ni/Mg,Si,Ca ratios consistent with the (equal) solar 
mix and \cr source ratios.
Basically, this amounts to showing that the current SN\,Ia and SN\,II models,
together with the estimated IMF, can, by and large, account for the
solar mix Fe,Ni/Mg,Si,Ca ratios.
This applies to the SNR grain material as well, since all these
elements are believed to be soon entirely locked in grains in SNR's.

Lingenfelter \etal therefore contend that the \cr Fe,Ni/Mg,Si,Ca ratios
are accounted for, if \crs are accelerated from SNR grain material.
{\it This is, however, true only if the relative contributions of the
various \SNIa and \SNII are equal, to within 20~\%, for \oi the
galactic enrichment of each of the various species, and \ii the
corresponding amount of \cr accelerated material.\/}
This requires that the \cr acceleration yields follow, without any
bias, the yields for the enrichment of the various elements for all \SNo.
While this is not impossible, it does not seem likely since these 
various objects, \SNIa and \SNII of all masses, have very different 
layer structures, ejection speeds, ejecta masses, and ISM 
environments.


\subsubsection{2.3.2. The main-s-process elements ($A$ \gappeq 87)}

\subsubsection{2.3.2.1. Main point !}

As discussed in {\S}~2.1, s-process species must be subdivided between
weak-s species with $A$ \lappeq 87, made in massive stars (no observed
GCR refractory), and main-s species with $A$ \gappeq 87, synthesized
in low mass, 1~--~3 \Msol stars during the AGB phase.
Here, we are interested in these main-s elements, for which we have a
sample of four refractory elements with determined GCRS abundance: Sr,
Zr, Ba, Ce.

These main-s elements are definitely {\it not\/} made, to a
significant fraction, in any type of SN ({\S}~2.1)~!
%
Nevertheless, they are {\it not\/} underabundant in GCRS's, relative 
to Mg, Al, Si, Ca, Fe, Ni, Pt-group, and Actinides, which are all made 
in \SN (Fig.~2).
So, we conclude that SN nucleosynthesis {\it cannot\/} control the
current GCRS composition (Prantzos \etal 1993).

\subsubsection{2.3.2.2. Observed Ba enrichment in SN 1987A ?}
Ba is an almost pure main-s element.
%
Now, Mazzali \etal (1992) have reported the observation of a Ba/Fe
enhancement by a factor of 3.7 in SN 1987A.
This has led Lingenfelter \etal (1998) to question the conventional
views on the origin of the main-s-process elements in low mass AGB
stars, and to suggest that they may largely originate in SN\,II \nso.
The {\it lack\/} of a relative deficiency of the main-s elements in
cosmic rays would then no longer conflict with a SN ejecta origin of
the cosmic ray material.
We think that this questioning of the low mass star origin of the
main-s elements is {\it not\/} justified, on three grounds:
\parano \oi First, the reality of the observed Ba enhancement is far
from certain.  It is based on an analysis of Ba II lines, whose atomic
physics is not at all well under control.  Earlier analysis had
actually led to Ba/Fe enhancements by factors of 10 to 20.  This
factor has been reduced to $\sim$~3.7 in Mazzali et al.'s study, due
to their consideration of line blocking, leading to a higher predicted
Ba II/Ba III line strength ratio.  The authors themselves are
remarkably prudent regarding their analysis, stating: ``We also cannot
rule out the possibility that other recombination mechanisms, which we
have not considered in this work, may be important as well.  The fact
that [\ldots\ldots] suggests that perhaps no real s-process elements
overabundance is present in SN 1987A".
\parano \ii Second, main-s-process \ns in \SNII, if significant,
would be expected to yield larger excesses of lighter elements.  In
particular, the $_{38}$Sr excess is expected to be significantly
larger than that of $_{56}$Ba (Prantzos \etal 1988).  Now, Mazzali et
al.'s analysis yields a Sr/Fe excess by a factor of $\sim$~1.5 only~!
This does not add credibility to the abundance analysis.
\parano \iii The evolution of the galactic Ba abundance over the range 
[Fe/H] = -- 2 to 0 shows beyond doubt that Ba is, indeed, 
predominantly produced by low mass stars (Edvardsson \etal 1993; 
Gratton \& Sneden 1994; McWilliam 1997; Mashonkina \etal 1999).
\parano We conclude that the existing Ba observations in SN 1987A do {\it
not\/} justify a questioning of the predominantly low mass star origin
of the main-s elements.

\subsection{The GCRS Isotope Ratios}

As well known, the GCRS \Nerat ratio is about 4.5 times solar, and this
suggests the presence of a He-burning material component in GCRS's,
probably originating in WC-type WR star wind material
({\S}~1.2).
An associated \Ct excess is expected, which is, indeed, suggested by
the recent analysis of the \Crat ratio (Duvernois \etal 1996; Webber
\etal 1996).

With this sole exception, the outcome of a large number of recent 
studies of the isotopic ratios
%
\footnote{References for GCRS isotope ratios: Leske 1993 (Fe,Co,Ni);
DuVernois \etal 1996 (C,N,O,Ne,Mg,Si); Westphal \etal 1996 (Fe,Ni);
Webber \etal 1996 (C,N,O),1997 (Ne,Mg,Si,S); Connell \& Simpson 1997
(Fe,Co,Ni); Lukasiak \etal 1997a (Co,Ni),1997b (Ca,Fe); George \&
Wiedenbeck 1998 (Cu,Zn).}
%
shows that the GCRS isotopic ratios of {\it all} other measured 
elements (N, O, Mg, Si, S, Ca, Fe, Co, Ni, Cu, Zn) are consistent with 
the solar mix
%
\footnote{Connell \& Simpson (1997) have reported source
\Feffouro\hsla\Fefsix and \Fefseveno\hsla\Fefsix enhancements by factors of
$\sim$~1.2 to 1.6 and of $\sim$~1.6, respectively.  The source
\Feffouro\hsla\Fefsix excess is sensitive to the interstellar propagation
calculation and to the \Mnffour lifetime.  Its high value is {\it
not\/} confirmed by the studies of Leske (1993) and Lukasiak \etal
(1997b).  As for the \Fefseveno\hsla\Fefsix ratio, mass 57 seems very
poorly resolved in Connell \& Simpson's data.}.


\subsection{The \Nifn Clock}                          

\subsubsection{2.5.1. General}
~ \vskip 0.3truecm

\parano
\Nifn is an unstable isotope which decays by electron capture with a
period \teco~=~$1.1~10^{5}$ yr, {\it provided that\/} it is not fully
stripped.
In the conventional view, the stable galactic \Cofn has been initially
synthesized as \Nifno.
So, if cosmic rays are accelerated out of fresh SN ejecta
$\lappeq 10^{5}$ yr after the explosive \nso, the initial \Nifn
nuclei have not had enough time to decay before they are fully
stripped by the acceleration, and are preserved in cosmic rays.
If, by contrast, cosmic rays originate in ``old" ISM (or even
circumstellar) material, the initial \Nifn nuclei have
had plenty of time to decay before their acceleration, and no \Nifn
should be observed in cosmic rays (Soutoul \etal 1978).

We have observations of the cosmic ray \Nifn abundance by Connell \&
Simpson (1997) and Lukasiak (1997a), which have been recently
outclassed by the new ACE data (Wiedenbeck \etal 1999), shown in
Fig.~3.
These data show that $<$~25~\% of the mass 59 material has been
accelerated in the form of the \Nifno.
So, it seems that  all, or most of, the \Nifn has decayed in the
cosmic ray source material, \ie that the acceleration took place
$\gappeq 10^{5}$ yr after the explosive \nso.  Since the typical time
for dilution of the SN ejecta is a few 10$^{4}$ yr, this implies that
cosmic rays do not originate in fresh ejecta material
%
\footnote{The relevance of the \Nifn test to \ns could, however, be
invalid, if supposedly accelerated \Nifn nuclei can pick up electrons
during their interstellar propagation, allowing them to decay ``en
route".  But the ACE data refer to the energy range 120 -- 600
MeV/nucleon.  In view of the roughly flat shape of the observed \cr
energy spectra in this range, a major fraction of the observed
particles lie towards the upper part of this range, where electron
pick up can be excluded (even if there exists a moderate amount of
interstellar re-acceleration).}.

%
%
\begin{figure}[h]                   
\resizebox{\hsize}{!}{\includegraphics{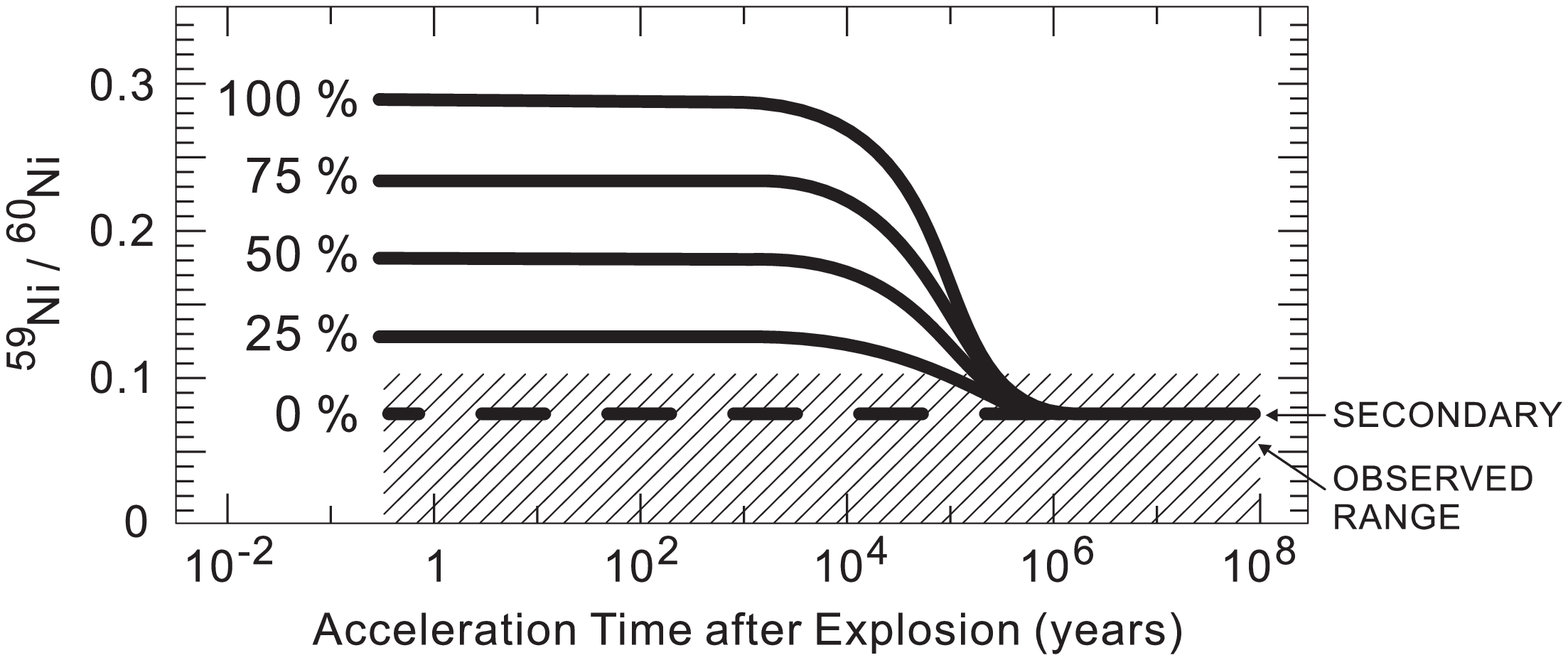}}
\caption{The \Nifno/\Nis ratio, as observed by the ACE/CRIS instrument
(upper limit; hatched), and as calculated as a function of the time
elapsed between the SN explosive \ns and the acceleration of the \cr
particles ({\S}~2.5).  The calculated curve are labeled by the
fraction of the mass 59 initially synthesized in the form of \Nifno;
the 0 \% level corresponds to the secondary production of \Nifno.
After Wiedenbeck \etal (1999).}
%
\end{figure}
%
%
%

\subsubsection{2.5.2. Discussion:  A direct \ns of \,$^{\it 59}$Co ?}
~ \vskip 0.3truecm

\parano
This is all true, provided that most of the \Cofn has, indeed, been 
first synthesized in the form of \Nifn in the \SN associated with 
\cr acceleration~!
Otherwise, the test on the time delay is not applicable.
With the ACE data, indicating that in the \cr sources 
\Nifno/(\Nifno+\Cofno) $<$ 0.25, it is actually sufficient that 
$\gg$~25~\% of the \Cofn has been synthesized in the form of \Nifn for 
the test to work.

Lingenfelter \etal (1998) have questioned that most of the \Cofn has
been first synthesized in the form of \Nifno.  Based on Woosley \&
Weaver (1995)'s models for SN\,II, they have estimated that as much as
$\sim$~50~\% of the \cr \Cofn has been directly synthesized as \Cofno.
We now reexamine this situation.

Actually, only $\sim$~30~\% of the galactic Fe,Co,Ni has been
synthesized in \SNIIo, and as much as $\sim$~70~\% in \SNIa
(see discussion in {\S}~2.3.1)
%
%
Let us now examine, in turn, the mass 59 production of \SNII and
\SNIao.
\vskip 0.1truecm
\parano \oi In \SNIIo, the e-process produces only \Nifn, essentially
{\it no\/} \Cofn (\egc Thielemann \etal 1996).
However, \Cofn can be produced directly by two processes: the pre-SN 
weak-s-process, and neutrino spallation (Woosley \& Weaver 1995; 
Chieffi \etal 1998).
While the SN\,II initial \Nifno/(\Nifno+\Cofno) ratio depends on the 
stellar mass and is very sensitive to the mass cut for massive stars, 
Lingenfelter \etal (1998) have performed an averaging of the Weaver \& 
Woosley yields over the IMF, and obtained an overall 
\Nifno/(\Nifno+\Cofno) ratio of $\sim$~0.50 for all \SNII
%
\footnote{In Thielemann \etal (1996), {\it no\/} \Cofn seems to be
made directly, because this study does {\it not\/} include the pre-SN
weak-s-process and the neutrino spallation.}.
%
%
%
\vskip 0.1truecm
\parano \ii In \SNIao, the e-process similarly produces only \Nifn,
essentially {\it no\/} \Cofn (\egc Thielemann \etal 1986).
But there exists {\it no\/} weak-s-process and neutrino spallation.
So, there is no direct \Cofn synthesis, and the \Nifno/(\Nifno+\Cofno)
ratio is 1 for \SNIao.
\vskip 0.1truecm

We conclude that, {\it even if\/} \SNII made {\it all\/} mass 59 in
the form of \Cofno, all \SNIa and II together would yield an initial
\Nifno/(\Nifno+\Cofno) of $\sim$~0.70.
This represents a lower limit.
With Lingenfelter et al.'s  estimate of an IMF-averaged
\Nifno/(\Nifno+\Cofno) $\sim$~0.50 for \SNII only, we would get a total
initial \Nifno/(\Nifno+\Cofno) of $\sim$~0.85.
So, in any case, the initial \Nifno/(\Nifno+\Cofno) is $>$~0.70, which
is much larger than the upper limit of 0.25 found for the \cr sources.

Thus, if the yields of \cr Fe, Co, and Ni from the various types of 
\SN are at all similar to those for galactic \nso, these 
considerations apply to \crso.  Then, the ACE data prove that the 
acceleration took place $\gappeq 10^{5}$ yr after the explosive \nso, 
and that {\it no\/} fresh SN ejecta are accelerated
%
\footnote{Quite extreme and contradictory assumptions would be
required for this conclusion {\it not\/} to be valid~!  It would
require that essentially {\it only\/} \SNII accelerate \crso, which we
have no reason to believe.
And even then, Lingenfelter et al's IMF-average over \SNII yields 
\Nifno/(\Nifno+\Cofno) ratios of $\sim$~0.50 $\gg$ 0.25.
So, for the \Nifn evidence {\it not\/} to be relevant, one would have to
require that \crs are accelerated specifically by those \SNII that
produce essentially no \Nifno, thus producing all their \Cofn directly.
A very far fetched hypothesis indeed~!
In Woosley \& Weaver's work, actually, this total lack of \Nifn
production happens in some of the massive star ($\geq$ 30 \Msolo)
models which, due to the mass cut, also produce {\it no\/} dominant
\Nifso, and hence eventually very little \Fefso, at variance with the
\cr Fe/Mg,Si ratio.}.

\subsection{Current \CRs and Acceleration of Fresh SN Ejecta ? \\
            Conclusions}                              

For three fundamental reasons, we conclude that fresh SN ejecta {\it
cannot} be a significant source of the current cosmic ray material:
\vskip 0.15truecm \parano
\oi The \cr source composition shows {\it no\/} anomaly related to
SN\break \nso.
\parano First, the \cr source FeNi/MgSiCa ratios have precisely the 
solar mix values; if \crs originate in SN ejecta, this requires that 
the efficiency of accelerating the ejected material is nearly 
identical for the various layers of any particular SN, as well as for 
the various \SNIa and \SNII of all masses; while not impossible, this 
does not seem likely ({\S}~2.3.1).
\parano Further, the main-s-process elements, which are not 
significantly synthesized in {\it any\/} type of SN, are {\it not\/} 
underabundant relative to all elements synthesized in \SNo; 
this~is~clearly inconsistent with a significant acceleration of fresh 
SN ejecta ({\S}~2.3.2).
\parano Finally, almost all \cr source isotopic ratios are consistent
with the solar mix.  There is one exception: a large \Nett and probably
\Ct excess, which is indicative of an acceleration of pure He-burning,
Wolf-Rayet star wind circumstellar material, {\it not\/} of SN ejecta
material ({\S}~1.2 and 2.4).
\vskip 0.15truecm \parano
\ii The absence of \Nifn in \crs implies that the time delay
between the SN \ns of Fe peak nuclei and their acceleration is
$\gappeq 10^{5}$ yr ({\S}~2.5).
\vskip 0.15truecm \parano
\iii The physics of SNR's and of \cr shock acceleration makes a
significant acceleration of {\it interior\/} ejecta material very
difficult; the acceleration of {\it external\/}, interstellar and/or
circumstellar material is expected to be largely dominant.  This is
discussed in the companion paper by Ellison \& Meyer (1999).

\section{A Conflict between the Linear Evolution of Be/H in the Early
         Galaxy, and the Absence of Fresh SN Ejecta in Current \CRs ?}

Do we have a conflict between
\oi the early Galactic evolution of Be/H, indicating that most of the
\cr CNO was {\it then\/} originating in freshly synthesized material, and
\ii the current \cr composition and acceleration conditions,
indicating that most {\it current\/} \crs do {\it not\/} originate in
fresh SN ejecta, but rather in interstellar and/or circumstellar
material
%
%
\footnote {With currently $\sim$~38~\% of the source, and $\sim$~33~\% 
of the propagating CNO originating in freshly synthesized WR wind 
material.  This fresh WR \,CNO is responsible for $\sim$~7~\% of the 
currently produced Be.  See footnote in {\S}~1.3.}.
%
%
%
Not necessarily~!  {\it \Crso , indeed, need not be the same now, and in
the early Galaxy~!\/}
We may also note that, for the current epoch, we know the \cr
composition, but not the rate of the Be/H evolution, while for the
early Galaxy, we know the rate of Be/H evolution, but not the \cr
composition~!

We will first ask ourselves whether the Be/H evolution actually {\it 
requires} a predominance of \crs accelerated out of fresh \ns 
products in the early Galaxy.
Then, we will very briefly explore possible sources for such freshly
synthesized \crs in the early Galaxy.

\subsection{Were the Be Producing \CRs Representative \\ of the Bulk of
the \CRs in the Early Galaxy ?}                      

First, is there really a contradiction ?

As discussed in {\S}~1.1.1, there was very little CNO in the ISM in the
early Galaxy, so that
almost {\it no\/} Be could be produced by the spallation of both ISM\e
CNO by \cr protons (Be$_{\rm ism}$, dominant in the current Be
production~!), and of \cr CNO {\it accelerated out of the ISM\/} on ISM
Hydrogen.
Only \crs accelerated out of freshly synthesized, locally CNO-enriched,
material could yield a significant contribution to the Be production.
{\it In earlier times, any such contribution to the Be production, 
however limited, must therefore have been dominant.\/}
It is thus not surprising to have Be evolve as a primary in the very
early Galaxy.

Did these CNO enriched \crs represent the bulk of the \crs in the early
Galaxy ?  We don't know.
There may have then existed many more \crs accelerated out of the then
metal-poor ISM, essentially composed of protons and \alpho-particles
only, which could not produce any Be.
The ``Be indicator" is blind to them.

We see two avenues to set upper limits to such an hypothetical
H,He-rich \cr component of interstellar origin: \oi the energetics,
and \ii the evolution of \Lisxo, which can be produced by \alpho-\alph
reactions.

\subsection{Possible Sources for Freshly Synthesized \CRs
            \\ in the Early Galaxy}                   

With the above considerations in mind, we now consider three possible
sources for freshly synthesized, CNO-rich \crs in the early Galaxy.

\subsubsection{3.2.1 Directly Accelerated SN Ejecta in the Early Galaxy}
~ \vskip 0.3truecm

\parano One may wonder if the direct acceleration, by each particular
SNR, of {\it even a small amount\/} of its own SN ejecta material
might have yielded, in the early Galaxy, a {\it comparatively\/}
significant contribution to the formation of Be.
Such an acceleration of some ejecta material can take place in
two ways:
\oi There will always be some acceleration by the reverse shock.  Even 
if this acceleration is small overall, it might be more efficient for 
producing Be than for producing \cr particles since, even if the 
accelerated particles are later adiabatically decelerated and never 
escape, the Be they have produced will survive.
In addition, both the Be created in flight, Be$_{\rm fast}$, and at
rest, Be$_{\rm rest}$, are created out of material highly enriched in
CNO, and {\it none\/} will escape the Galaxy (Parizot \& Drury 1999).
\ii The forward shock may be overcome by clumps of fast-moving
ejecta, which get later accelerated as "external" material (Jun \&
Norman 1996).  This, however, should concern only a small part of the
total ejecta mass.

On the other hand, both phenomena can occur only in the very early
phases of the SNR lifetime.
Further, the {\it total\/} ejecta mass is but $\sim$~2 \x{-3} to
3 \x{-2} of the total ISM mass swept by the forward shock, and this is
true in the early Galaxy as well as to-day (Drury \& Keane 1995;
Parizot \& Drury 1999; Ellison \& Meyer 1999).
All in all, the acceleration of their own ejecta by SNR's seems
to be a comparatively small phenomenon, even in the early Galaxy
(Parizot \& Drury 1999).

\subsubsection{3.2.2 Wolf-Rayet's in the Early Galaxy ?}
~ \vskip 0.3truecm

\parano As discussed in {\S}~1.2 and 1.3, a significant fraction
($\sim$~33~\%) of the {\it current\/} propagating \cr CNO is primary,
originating in WR star wind material.  It is responsible for some
$\sim$~7~\% of the currently produced Be.
One may, of course, wonder: could this component have been larger in
the early Galaxy ?

The answer is a definite: no, there were fewer WR's than to-day~!
As compared to O-stars, the observed number of WR stars, indeed,
strongly decreases for decreasing metallicity (\egc Maeder \& Conti
1994; Maeder \& Meynet 1994)~\nobreak
%
\footnote{The basic interpretation for
this decrease is that the huge WR winds are primarily driven by the
radiation pressure exerted on the heavy elements in the outer stellar
layers.
Since, in the early Galaxy, there were essentially no heavy elements
in the outer, {\it un-processed\/}, layers, the stellar peeling off
process could not be initiated.
So, it was difficult to produce  single WR's in the early Galaxy.
More precisely, the stellar mass threshold for the onset of the WR
phenomenon increases with decreasing metallicity.
Other factors probably play a role.  Binarity can probably induce WR
winds, although there is no clear evidence for a larger fraction of
binaries among WR's in metal-poor galaxies (SMC) (\egc Moffat 1995),
and the importance of the role of binarity in initiating the WR
phenomenon is currently controversial (Maeder \& Meynet 1994; Maeder
\& Conti 1994; Vanbeveren \etal 1997,1998).  Rotational mixing of
massive stars layers may also play a role in enriching the outer
layers in heavy elements.}.
%
In {\it relative\/} terms -- as compared to \crs accelerated out of
the then CNO-poor ISM -- , however, the WR component could have played
a significant role for the Be production in the early Galaxy
%
\footnote{Note, however, that this possible early Galactic WR
production of Be refers to the Be$_{\rm fast}$ component only
({\S}~1.1), which {\it currently\/} accounts for only $\sim$~20~\% of
the forming Be ({\S}~1.3).  The same remark applies to the contribution of
the \SN exploding within \sbso, discussed just below ({\S}~3.2.3; see 
also footnote in {\S}~1.1.1).}.
%

\subsubsection{3.2.3 \SN Exploding Within Superbubbles
                     in the Early Galaxy} ~ \vskip 0.3truecm

\parano Another way to have an acceleration of material enriched in
freshly synthesized heavy elements is to consider an OB association
forming a superbubble, within which some of the ambient material has
been locally enriched by the WR winds and the ejecta of the previously
exploded \SNo.
This ambient superbubble material can be accelerated, both by the
expanding forward shock waves of individual SNR's, and by a general
turbulence developing within the superbubble medium.
This possibility has been considered  by a number of authors recently
(Bykov 1995,1999; Parizot \etal 1998; Parizot 1998; Vangioni-Flam 
\etal 1998a; Higdon \etal 1998; Parizot \& Drury 1999; Ellison \& 
Meyer 1999)
%
\footnote{Sometimes in connection with the high nuclear \gammo-ray
fluxes earlier reported for the Orion nebula,  which have {\it not\/}
been confirmed since then (Bloemen 1999).  These data implied large
accelerated particle fluxes, strictly limited to low energies.
Associated theoretical work referenced in the above papers.}.

Clearly, such a \sb component, enriched in fresh \ns products, cannot
be dominant among {\it current\/} \crso.
Difficulties encountered by this hypothesis, in terms of the \sb and the
acceleration physics, are discussed in the companion paper by
Ellison \& Meyer (1999).
Regarding composition, arguments similar to those developed in {\S}~2
against a predominance of SN ejecta among current \crso, also apply
here.
Essentially, OB associations contain  only \SNIIo, and no \SNIao.
Now, the nearly solar value of the \cr source Fe,Ni/Mg,Si,Ca ratios is
clearly inconsistent with the acceleration out of a medium enriched by
massive \SNII only, with no \SNIa contribution.  Roughly, one would
then expect a 3-fold relative deficiency of Fe,Ni ({\S}~2.3.1).
The lack of a relative main-s-process element deficiency is also
inconsistent with a significant acceleration of current \crs out of
such a medium ({\S}~2.3.2).

Could \sbs have played a more important role in the generation of
\cr CNO nuclei in the early Galaxy~?
The possibility that a larger fraction of massive stars might have
formed within large OB associations in the gas-rich, very active,
early Galaxy could be explored.
But in any case, in {\it relative\/} terms -- as compared to \crs
accelerated out of the then CNO-poor ISM -- such a \sb contribution could
have  played a significant role for the Be production in the early
Galaxy.

\acknowledgments{We wish to thank Marcel Arnould, Michel 
Cass\'e,\break Georges Meynet, Robert Moscovitch, Nikos Prantzos, 
Hubert Reeves, Elisabeth Vangioni-Flam, and Andrew Westphal for 
numerous discussions, with have been important for the elaboration of 
this paper.
This work was supported, in part, by NASA's Space Physics Theory
Program.}



\end{document}